\documentclass[twocolumn]{aastex62}
\usepackage{float,graphicx,amsmath,multirow,mathtools}
\usepackage{color}
\usepackage[version=4]{mhchem}
\usepackage{comment}
\usepackage{amsmath}
\usepackage[caption=false]{subfig}
\usepackage{comment}

\usepackage{bm}
\newcommand{\mjb}{mJy~beam$^{-1}$}
\newcommand{\kms}{km~s$^{-1}$}

\newcommand{\msunyr}{M$_{\Sun}$\ yr$^{-1}$}
\newcommand{\msun}{M$_{\Sun}$}

\newcommand{\cmnegthree}{cm$^{-3}$}

\newcommand{\Rdust}{$R_\mathrm{dust}$}
\newcommand{\Tcomp}{$T_\mathrm{comp}$}
\newcommand{\Rcomp}{$R_\mathrm{comp}$}
\newcommand{\fvol}{$f_\mathrm{vol}$}
\newcommand{\fic}{$f_\mathrm{ic}$}

\begin{document}

\title{Investigating Anomalous Photochemistry in the Inner Wind of IRC+10216 Through Interferometric Observations of \ce{HC3N}}
\author{Mark A. Siebert}
\affiliation{Department of Astronomy, University of Virginia, Charlottesville, VA 22904, USA}
\author{Marie Van de Sande}
\affiliation{School of Physics and Astronomy, University of Leeds, Leeds, UK}
\author{Thomas J. Millar}
\affiliation{School of Mathematics and Physics, Queen’s University Belfast, Belfast, UK}
\author{Anthony J. Remijan}
\affiliation{National Radio Astronomy Observatory, Charlottesville, VA 22903, USA}

\submitjournal{ApJ}
\accepted{October 26, 2022}

\begin{abstract}
In recent years, many questions have arisen regarding the chemistry of photochemical products in the carbon-rich winds of evolved stars. To address them, it is imperative to constrain the distributions of such species through high angular resolution interferometric observations covering multiple rotational transitions. We used archival ALMA observations to map rotational lines involving high energy levels of cyanoacetylene (\ce{HC3N}) toward the inner envelope (radius $<8"$/$1000$\,AU) of the carbon star IRC+10216. The observed lines include the $J=28-27$, $J=30-29$, and $J=38-37$, transitions of \ce{HC3N} in its ground vibrational state. In contrast to previous observations of linear carbon chains toward this AGB star which show extended, hollow emission at $15"-20"$ radii (e.g.\ \ce{C4H}, \ce{C6H}, \ce{HC5N}), the maps of the \ce{HC3N} lines here show compact morphologies comprising various arcs and density enhancements, with significant emission from gas clumps at an angular distance of ${\sim}3”$ (350\,AU) from the central AGB star. We compared visibility sampled non-LTE radiative transfer models with the observed brightness distributions, and derive a fractional abundance with respect to \ce{H2} of $10^{-8}$ for \ce{HC3N} at the radii probed by these lines. These results are consistent with enhanced photochemistry occurring in warm (${\sim}$200 K) regions of the circumstellar envelope. After application of a specialized chemical model for IRC+10216, we find evidence that the enhanced \ce{HC3N} abundances in the inner wind are most likely due to a solar-type binary companion initiating photochemistry in this region.

\end{abstract}
\keywords{Astrochemistry -- circumstellar matter -- line: identification -- stars: AGB, individual (IRC+10216)}

\section{Introduction}
IRC+10216, also known as CW Leo, is a well-studied nearby \citep[$d\sim130$\,pc,][]{Menshchikov2001_IRCphys} Asymptotic Giant Branch (AGB) star with a chemically rich expanding circumstellar envelope (CSE). The CSE is the result of dust-driven mass loss at a rate of ${\sim}2\times10^{-5}$\,\msunyr \citep{DeBeck2012_IRCphysCO}, and represents one of the final stages of evolution for 1--8\,\msun\,stars immediately preceding the transition between the AGB and the formation of a planetary nebula (PN). During this stage, depending on how much carbon has been convectively dredged up from the star's core relative to oxygen \citep{Weiss2009_AGBmodels}, AGB stars can either be M-type (C/O$<1$), C-type (C/O$>1$), or S-type (C/O${\sim}1$). IRC+10216 has a C/O ratio of 1.4 \citep{Winters1994_CSEcurves}, which allows for a complex system of carbonaceous gas phase and dust grain chemistry to occur throughout its extended envelope. To date, more than 95 unique molecules have been identified toward IRC+10216 \citep{McGuire2022_census}. Due to its close proximity and history as a molecular line object, IRC+10216 forms the basis of our understanding on how carbon-rich stellar material is processed and returned to the interstellar medium by low-to-intermediate mass stars.

An important aspect of AGB evolution that has become clear with the advent of ALMA is the prevalence of binary companion stars influencing mass loss and shaping circumstellar winds \citep{Decin2021_massloss_review}. Companions are difficult to detect directly in these systems, due to the intrinsic luminosity of the AGB star, and the high attenuation of visible light by their dusty envelopes. However, several recent studies of gas and dust kinematics have shown the perturbative effects binary companions can have on many AGB and proto-planetary nebula (PPNe) systems. Both small-scale asymmetries, such as density-enhanced clumps \citep[e.g.][]{Khouri2016_vibCO_AGBs, Agundez2017_IRC_carbonchains, Leao2006_IRC_highresOptical}, and large-scale structures, such as spirals \citep[e.g.][]{MauronHuggins2006_CSE_Imaging,Maercker2016_RScl_shell, Decin2015_IRC_spiral, Ramstedt2017_WAql_binary}, disk geometries \citep[e.g][]{Kervella2014_L2Pup_disk, Homan2018_RDor_Disk, Sahai2022_VHyaDUDE, Homan2020_piGru_atomium}, and bipolar outflows \citep[e.g.][]{Kim2015_CIT6Bipolar,Sahai2017_IRAS05506_bullets, Lagadec2011_IR_pAGBs} have been detected. In particular, the ALMA Large Program ATOMIUM (2018.1.000659.L, PI: Decin) has shown the widespread nature of these structures through a survey of numerous AGB outflows \citep{Decin2020_atomium}. It is now thought that influence from stellar and planetary companions is closely related to the stark morphological changes in circumstellar material during the transition between the AGB and PN stages of evolution \citep{Sahai2007_PPNsurvey_class}. IRC+10216 shows signs of binary influence, based on the interspaced shells in its extended envelope \citep{cernicharo2015_IRC_MLRhist,Guelin2018_IRC_3D_Morph}, spiral structure in its inner layers \citep{Decin2015_IRC_spiral}, as well as the detection of a tentative compact source $0.5$\arcsec\ from the star via archival HST monitoring by \cite{Kim2015_IRC_companion_HST,Kim2021_IRC_HSTmonitor2}.

While their physical effects have been studied in detail, the chemical impact of companion stars embedded in CSEs and PPNe has not yet been observationally tested. The typical framework for how gas phase chemistry proceeds in circumstellar envelopes assumes precursor molecules (e.g. \ce{CO}, \ce{HCN}, \ce{SiO}) form out of a combination of equilibrium and shock-induced chemistry in the high temperature regions near the stellar photosphere \citep{Agundez_AGB_TE_Review_2020,Cherchneff_IRC_Shocks2012}. Once they reach the outer regions of the envelope where the density of attenuating dust grains is lower, UV photons from the interstellar radiation field (ISRF) initiate a kinetically-driven photochemistry to form a variety of exotic product molecules (e.g.\ \ce{HC7N}, \ce{SiN}, \ce{MgNC}) \citep{Ziurys2006_CSEchem, Li2016_AGBchemdist,Millar2020_UVinCSE}. While this scheme is consistent with the measured abundances of many molecules, recent studies of IRC+10216 have revealed that this picture may be much more complicated than previously thought. \cite{Agundez2015_CH3CNmaps_IRC+10216} and \cite{QL2017_NaCN_IRC} discovered that \ce{CH3CN} and \ce{NaCN} are present in the inner envelope ($<5$\arcsec). This was an unexpected result as the proposed formation routes of both these molecules require photochemistry, which typically takes hold further out (10\arcsec--25\arcsec) in the CSE \citep{DinhVTrung2008_IRC_CPshells}. Similar results were also found for the product molecules water (\ce{H2O}) and diacetylene (\ce{C4H2}) \citep{Agundez2010_H2O_CSE_photo,Fonfria2018_IRC_C4H2}

One molecule that is central to photochemistry in CSE environments is cyanoacetylene (\ce{HC3N}). In the gas phase, this species is formed primarily through the reaction between acetylene (\ce{C2H2}) and the cyanide radical (\ce{CN}), the latter of which is only available as a reactant when the flux of UV photons is large enough to efficiently dissociate the precursor molecule \ce{HCN} \citep{Agundez2017_IRC_carbonchains}. Because \ce{C2H2} is an abundant precursor molecule and one of the main carriers of carbon in C-rich CSEs \citep{Santoro2020_Cchem_AGBs}, \ce{HC3N} is one of the first product molecules to form out of UV-driven chemistry, making it useful in probing the photoprocessing of gas phase material in these objects.

\ce{HC3N} and longer cyanopolyynes in the envelope of IRC+10216 were studied with the VLA by \cite{DinhVTrung2008_IRC_CPshells} and subsequently with ALMA by \cite{Agundez2017_IRC_carbonchains}, where it was seen between 10\arcsec--20\arcsec (1300--2600\,AU). This is where the \ce{HC3N} abundance is expected to peak based on photochemical models of this CSE; however, these observations targeted low-lying rotational transitions (including states up to $J=12$ and $E_{\mathrm{up}}=34$\,K) which naturally trace lower temperature regions. In order to gain a more complete view of this molecule in the envelope of IRC+10216, and to better understand the onset of photochemistry in these environments, it is important to spatially constrain higher energy transitions of \ce{HC3N} that could be discernable in the warmer regions where \ce{CH3CN} and \ce{NaCN} were observed.

To this end, we present an archival study of IRC+10216 with the goal of characterizing \ce{HC3N} in the inner (arcsecond scale) layers of its envelope. We utilize observations from three separate ALMA projects to analyze the spectra and spatial distributions of the $J = 28 - 27$, $J = 30 - 29$, and $J = 38 - 37$ lines of \ce{HC3N} toward IRC+10216. Using maps of these lines at the systemic velocity combined with radiative transfer models, we quantify the abundance of cyanoacetylene in the inner regions of IRC+10216 and discuss how they affect our understanding of photochemical processing in this CSE. In Section \ref{observations} we describe the observations and reduction procedure, in Section \ref{maps} we present emission maps, in Section \ref{RTmodels} we use non-LTE radiative transfer models to derive the abundance profile of \ce{HC3N}, and in Section \ref{disc} we utilize a specialized photochemistry model to compare our observations with kinetic theory and test the hypothesis that molecules surrounding IRC+10216 are affected by a binary stellar companion embedded in the inner envelope.

\section{Observations and reduction}
\label{observations}

\begin{deluxetable*}{cccccccccc}
    \tablecaption{Summary and characteristics of the ALMA data utilized in this work. }
    \tablewidth{\columnwidth}
    \tablehead{
    \colhead{ALMA Project Code} & \colhead{Obs.\ Date(s)} & \colhead{Array(s)}  &\colhead{Synthesized beam}  & &\colhead{\ce{HC3N} line} & \colhead{Rest Frequency} & \colhead{$E_{\mathrm{up}}$} &\colhead{Image RMS}\\  & & & ($\theta_{\mathrm{maj}}\times\theta_{\mathrm{min}}$) &  & ($J'\rightarrow J''$) & (MHz) & (K) & (mJy/beam)} 
    \startdata
	2019.1.00507.S & Nov.\ 2019 -- Mar.\ 2020 & 12m+7m & $0.480"\times0.449"$ &  & 
	$28\rightarrow27$ & 254\,699.5 & 177 & 1.8 \\
	2011.0.00229.S & Apr.\ 2012 & 12m & $0.725"\times0.520"$ &  & 
	$30\rightarrow29$ & 272\,884.7 & 203 & 7.2 \\
    2016.1.00251.S & May 2018 & 12m & $0.848"\times0.761"$ &  & 
    $38\rightarrow37$ & 345\,609.0 & 324 & 1.9 \\
    \enddata   
    \tablecomments{Observations from ALMA Project 2019.1.00507.S include two 12m configurations and two from the 7m ACA. Transition frequencies and upper state energies were obtained from Cologne Database of Molecular Spectroscopy (CDMS) \citep{2005JMoSt.742..215M}. Reported RMS values were measured from an emission-free channel of the final reconstructed image cube.}
    \label{table:obspar-ALMA}
\end{deluxetable*}

\begin{figure*}[t!]
    \centering
    \includegraphics[width=\linewidth]{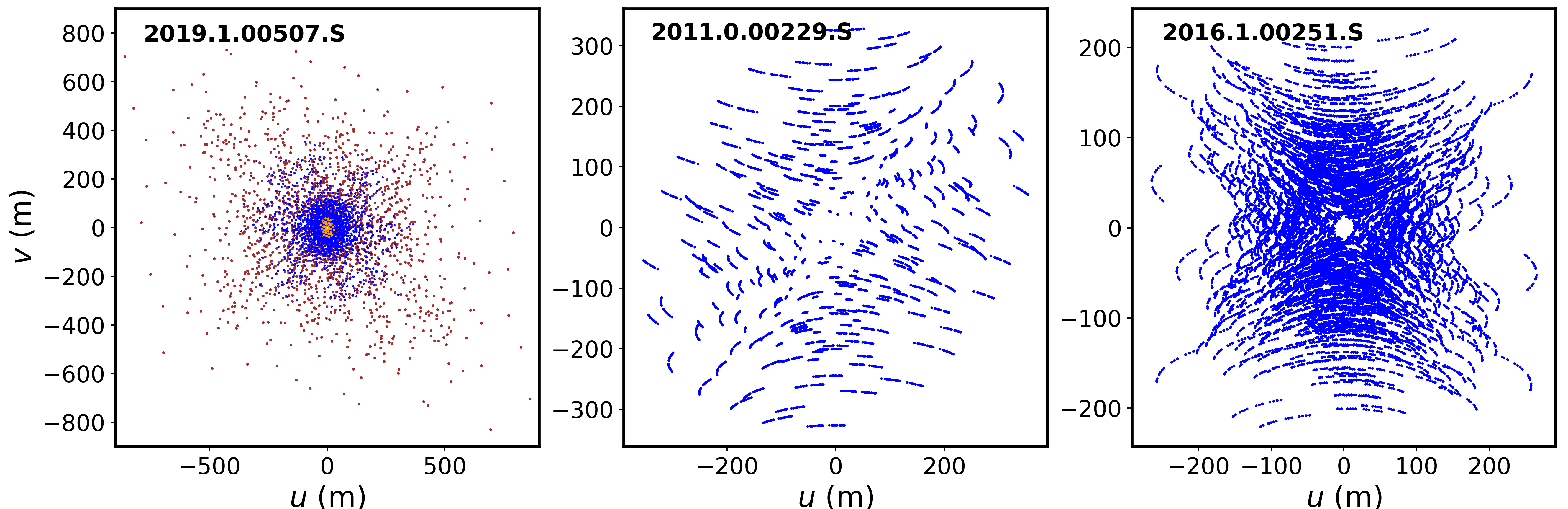}
    \caption{Coverage of the \textit{uv}-plane offered by the three ALMA projects utilized in this work. Baselines from compact 12m configurations are shown in blue, those from extended 12m configurations are shown in brown, and 7m ACA baselines are shown in orange. Observations from left to right were used to image the $J = 28 - 27$, $J = 30 - 29$, and $J = 38 - 37$ transitions of \ce{HC3N}, respectively.}
    \label{fig:uv_cov}
\end{figure*}

We utilize observations from three publicly available archival ALMA projects to map warm \ce{HC3N} emission toward IRC+10216. Fig.\ \ref{fig:uv_cov} demonstrates the different antenna configurations and coverage of the \textit{uv}-plane for the data sets, while Table \ref{table:obspar-ALMA} summarizes these different observations and the target \ce{HC3N} transitions for this work. Project 2019.1.00507.S provides several spectral scans of IRC+10216 at Band 6 between 250 and 270\,GHz using both the main 12m array and the 7m Atacama Compact Array (ACA). These observations were used by \cite{He2019_ALMAIRC_mmlinemonitoring} to study the time-variability of millimeter lines toward IRC+10216, and cover the $J=28-27$ transition of \ce{HC3N}. In order to recover flux from as many spatial scales as possible, we combined visibility measurements from four separate array configurations included in this project, including two 12m executions with baseline lengths ranging from 15\,m to 1230\,m and two ACA measurements covering baselines between 9\, and 50\,m (see Fig.\ \ref{fig:uv_cov}). Data combination was performed with antenna-specific weighting, and the resulting images have a maximum recoverable scale of $30$\arcsec.

Project 2011.0.00229.S provides a Cycle 0 spectral line survey of IRC+10216 covering most of Band 6 using the 12m array, allowing emission from the $J=30-29$ transition to be mapped as well; however, only 18 antennas were available at this time, as opposed to the 42 that were used for the later data sets. These data cover baselines ranging from 27\,m to 402\,m with a maximum recoverable scale of ${\sim}1"$.

\begin{figure*}[t!]
    \centering
    \includegraphics[width=\linewidth]{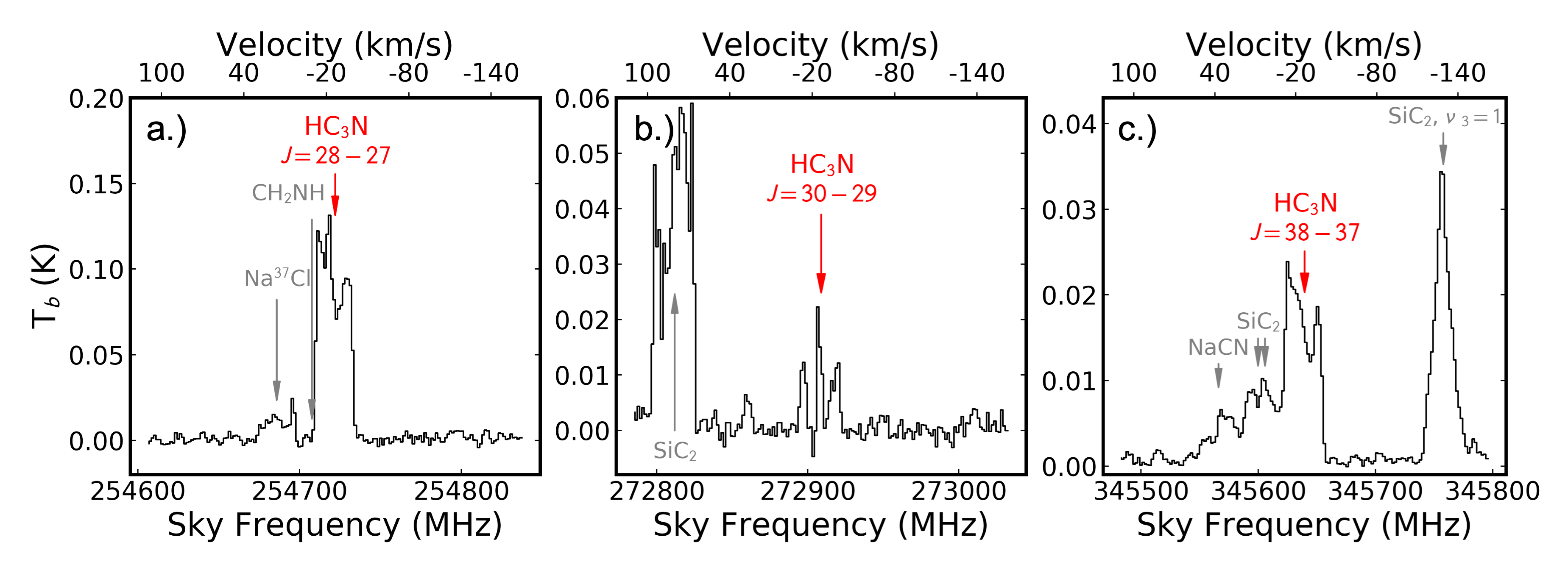}
    \caption{Spectra demonstrating the three \ce{HC3N} lines studied in this work (red), in addition to nearby and interloping molecular lines (gray). Spectra were created from image cubes using a 30\arcsec aperture centered on the position of IRC+10216. The lower x-axes denote sky frequencies, while the upper axis shows velocity using the rest frequency of the respective \ce{HC3N} transition. All transitions are from the ground vibrational state unless indicated otherwise.}
    \label{fig:specs}
\end{figure*}

Project 2016.1.00251.S provides a Band 7 observation between 330 and 346\,GHz using the 12m array. This ALMA configuration includes baselines between 15\,m and 314\,m, and offers a maximum recoverable scale of $8$\arcsec. These data were used to map the $J = 38 - 37$ line of \ce{HC3N} in addition to the $J = 3 - 2$ line of \ce{^{13}CO} at $\nu_{\mathrm{rest}}=330\,588.0$\,MHz. The \ce{^{13}CO} observations are useful in qualitatively characterizing the overall envelope density structure.

As Figure \ref{fig:uv_cov} demonstrates, the ALMA Projects we utilize have very different antenna setups, and consequently disparate spatial frequencies sampled. Due to these differences between instrument configurations and the non-uniqueness of the image reconstruction process, care must be taken when directly comparing brightness distributions between data sets. All observations were pointed near the continuum peak of IRC+10216 at $\alpha_\mathrm{J2000}$ = 09\fh47\fm57.458\fs, $\delta_\mathrm{J2000}$ = +13\arcdeg16\arcmin43.90\arcsec, despite having slightly different phase centers between projects. Errors in pointing are estimated to be 30\,mas for the $J=28-27$ observations, and 40\,mas for the $J=30-29$ and $J=38-37$ observations\footnote{Referencing Section 10.5.2\ of the ALMA Technical Handbook \citep{Cortes2020_ALMAtechnical}}.

All data were calibrated using the standard calibration pipeline of the Common Astronomy Software Application (CASA) \citep{McMullin2007_CASA}. Visibilities from the Band 6 data sets were resampled to a spectral resolution of 1.7\,\kms\ to match the correlator configuration in the Band 7 project. Continuum subtraction was done in visibility space using line free channels present in each spectral window containing \ce{HC3N}. Then, imaging was performed with the TCLEAN casatask, using Briggs weighting with a 0.5 robust parameter for the $J = 28 - 27$ and $J = 38 - 37$ lines, and natural weighting for the $J = 30 - 29$ line due to its lower signal-to-noise. Iterative masking was done with user-defined regions to optimize the deconvolution process. In the case of the $J = 28 - 27$ line, multi-scale cleaning was applied to account for the multiple combined array configurations \citep{Cornwell2008_multiscale_cln}. All rest frequencies were obtained from Cologne Database of Molecular Spectroscopy (CDMS) \citep{2005JMoSt.742..215M}, and shifted to the systemic velocity of IRC+10216, $V_{sys}=-26.5$\,\kms \citep{Cernicharo2000_IRC2mmSurvey}.

\section{Results}
\subsection{Spectra, interloping lines, and flux recovery}
\label{spectra}

Integrated spectra from each synthesized image cube were obtained using a 30\arcsec circular aperture centered on IRC+10216. These spectra are shown in Figures \ref{fig:specs}a-c. In these, we see emission in the characteristic ``double-horned" line profile centered on the systemic velocity of $-26.5$\,\kms, as well as some contributions from interloping lines. For the $J = 28 - 27$ line of \ce{HC3N}, a very close blend is seen with the $J(K_a,K_c)=4(0,4)-3(0,3)$ transition of methanimine (\ce{CH2NH}) at $\nu_{\mathrm{rest}}=254\,685.1$\,MHz. This line appears 14\,MHz (17\,\kms) from the central frequency of \ce{HC3N}. In the spectrum  (Fig.\ \ref{fig:specs}a.), a double horn profile of \ce{CH2NH} can be seen with the blue-shifted peak falling very close to the central channel of \ce{HC3N}. This blend was also detected by \citet{Tenenbaum2010_ARO_IRCsurv}. Since the aim of this study is to characterize the radial abundance using channel maps made at the systemic velocity, we must take care to make sure we do not attribute blue-shifted \ce{CH2NH} emission to \ce{HC3N}. However, considering the geometry of spherical outflows and empirical evidence from previous ALMA studies of AGB stars (e.g.\ channel maps in \cite{Decin2018_highlowMLR_alma} and \cite{Brunner2018_WAql_mols}), we expect that the contamination from the highest velocity emission is limited only to the smallest spatial scales and appears very close to the position of IRC+10216.

In the case of the $J = 38 - 37$ line of \ce{HC3N} (Fig.\ \ref{fig:specs}c.), there is a slight blend with the $J(K_a,K_c)=15(14,1)-15(14,0)$ transition of \ce{SiC2} at $\nu_{\mathrm{rest}}=345\,575.2$\,MHz. This lies 34\,MHz (29\,\kms) away from the central frequency of our target \ce{HC3N} line, so with a half line width of $14.5$\,\kms, only the most red-shifted channels of the image cube are contaminated by emission from \ce{SiC2}.

The $J=28-27$ and $J=30-29$ lines of \ce{HC3N} were also observed in the single dish line surveys of \cite{He_IRC_Survey_2008} and \cite{Tenenbaum2010_ARO_IRCsurv}, both using the Arizona Radio Observatory 10\,m Submillimeter Telescope (SMT). These works provide useful data points that allow us to characterize the amount of spatial filtering in our interferometric measurements of the same lines. Both \cite{He_IRC_Survey_2008} and \cite{Tenenbaum2010_ARO_IRCsurv} detected the $J=28-27$ line of \ce{HC3N} with a peak brightness temperature $T_R=130$\,mK. Considering that the SMT has a HPBW of 30\arcsec at 254.7\,GHz, we can compare this directly with the spectrum in Fig.\ \ref{fig:specs}a., which peaks at 120\,mK. In contrast, the $J=30-29$ line shown in Fig.\ \ref{fig:specs}b.\ was detected in \cite{Tenenbaum2010_ARO_IRCsurv} with a peak temperature of $T_R=90$\,mK, while we instead obtain a value of 10\,mK. It is clear that the observations from Project 2019.1.00507.S, which comprise both ACA and multiple 12m executions spanning a large range of baselines, are able to recover more than $90\%$ of the \ce{HC3N} ($J=28-27$) emission. However the Cycle 0 observations of the $J=30-29$ line exhibit a large amount of spatial filtering due to the fewer number of antennas and poorer \textit{uv}-coverage. Unfortunately, there exist no published single-dish measurements for the $J=38-37$ transition of \ce{HC3N} toward IRC+10216, but we still take steps to account for the flux loss of these measurements described in Section \ref{RTmodels}.
\subsection{Emission Maps}
\label{maps}
\begin{figure}[t!]
    \centering
    \includegraphics[width=0.93\linewidth]{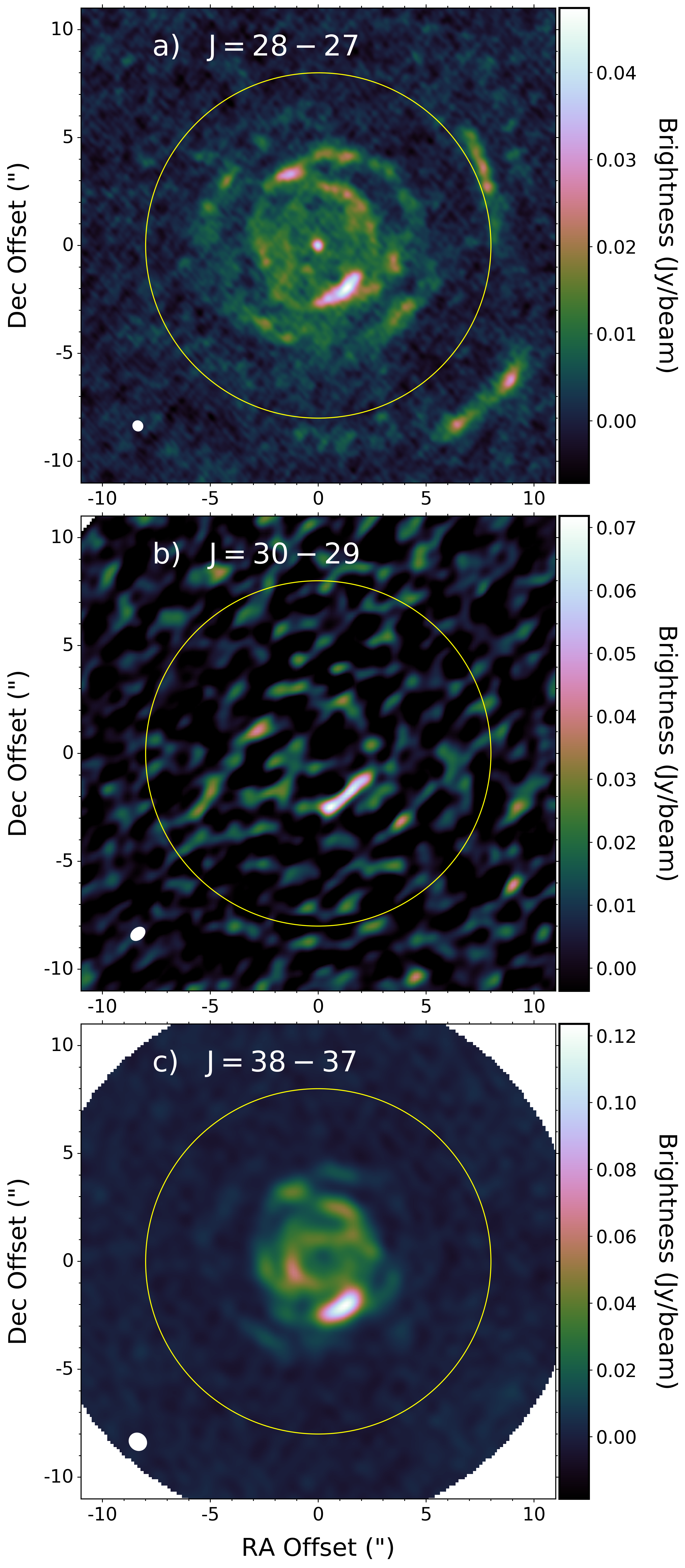}
    \caption{ALMA emission maps of three \ce{HC3N} lines toward IRC+10216. Images were created at the systemic velocity of -26.5\,\kms with a channel width of 1.69\,\kms. IRC+10216 is located at the center of the maps at $\alpha_\mathrm{J2000}$ = 09\fh47\fm57.458\fs, $\delta_\mathrm{J2000}$ = +13\arcdeg16\arcmin43.90\arcsec, and a reference circle with diameter 15\arcsec is shown in yellow. Because data from separate observation sets are used, the beam sizes and absolute brightness scales are different for each panel. The shapes of the synthesized beams are shown in the bottom left corner, with sizes 0.48\arcsec$\times$0.45\arcsec, 0.73\arcsec$\times$0.52\arcsec, and 0.85\arcsec$\times$0.76\arcsec from top to bottom.}
    \label{fig:tripfig1}
\end{figure}

Reconstructed maps of the three studied \ce{HC3N} lines are shown in Figure \ref{fig:tripfig1}. All images were made at the systemic velocity of IRC+10216 ($V_{sys}=-26.5$\,\kms). These maps show morphologies centered on the position of the AGB star with emission primarily within a radius of $5$\arcsec. The distribution of the $J = 28 - 27$ line includes an extended component out to 3\arcsec surrounded by several asymmetric arcs and clumps (Fig.\ \ref{fig:tripfig1}a.). Due to the higher temperatures probed by this transition, the majority of its emission traces regions entirely separated from that observed in the lower excitation lines mapped by \cite{Agundez2017_IRC_carbonchains} and \cite{DinhVTrung2008_IRC_CPshells} ($J=10-9$ and $J=5-4$; which peak at radius $14"$), with the notable exception of a region to the SW at a radius of $10"$. In the central component (within $3"$), the average brightness of the $J = 28 - 27$ line is 10\,\mjb, with notable bright clumps to the N and SW. Although a central point source is seen at the position of the carbon star, we expect that this is contributed from the most blue-shifted channel of the weak interloping line of \ce{CH2NH} (see Section \ref{spectra} and Fig.\ \ref{fig:specs}), since this peak is notably absent from the unblended image in Fig.\ \ref{fig:tripfig1}c. 

\begin{figure*}[t!]
    \centering
    \includegraphics[width=\linewidth]{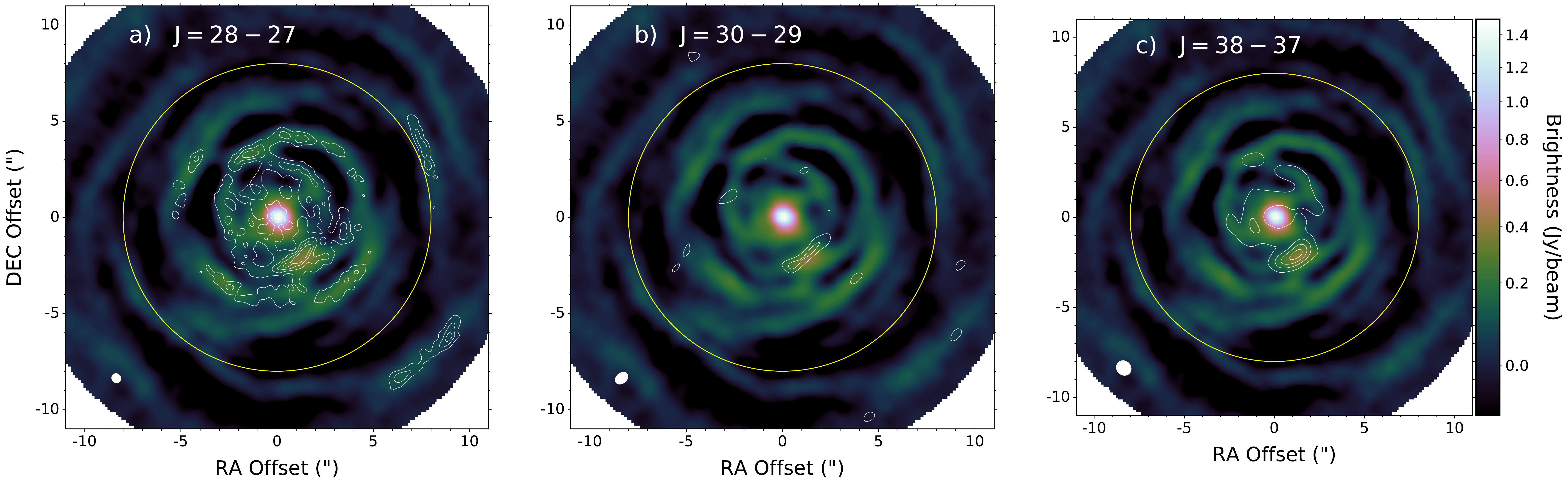}
    \caption{Contour maps of \ce{HC3N} $J=28-27$ $J=30-29$, and $J=38-37$ (white) overlaid on the image of \ce{^{13}CO} $J=3-2$ toward IRC+10216. All components are again taken at the systemic velocity and centered at the position of IRC+10216 $\alpha_\mathrm{J2000}$ = 09\fh47\fm57.458\fs, $\delta_\mathrm{J2000}$ = +13\arcdeg16\arcmin43.90\arcsec, and the yellow circle denotes a radius of 7.5\arcsec. Contour levels for each line are shown at 3, 6, 9, and 12 times the rms.}
    \label{fig:co_contours}
\end{figure*}

The $J = 38 - 37$ line (Fig.\ \ref{fig:tripfig1}c.) also shows bright, spatially resolved emission that is entirely distinct from the maps of \ce{HC3N} in \cite{Agundez2017_IRC_carbonchains}. In contrast to $J = 28 - 27$, the distribution of this line is entirely concentrated within a radius of $4"$ and shows no central peak on the position of IRC+10216, since there is no contaminating emission present in this channel. The imaged line is again not azimuthally symmetric, and traces the same bright clumps and arcs seen in the $J=28-27$ line. It reaches a peak brightness of $122\pm1.9$\,\mjb\ in the SW clump at $2"$. The more compact morphology can be qualitatively explained by the higher upper state energy (323\,K) and larger (by a factor of 3) Einstein $A_{ij}$ coefficient of this transition relative to the $J = 28 - 27$ line.

Finally, the Fig.\ \ref{fig:tripfig1}b.\ shows the $J = 30 - 29$ line, which also appears toward the same bright clump ${\sim}2$\arcsec to the SW. However due to the sensitivity of these data, very little additional emission is seen from this transition. If this line has a similar extended component as seen in $J = 28 - 27$, we would expect most of this flux to be resolved out by the interferometer given its maximum recoverable scale. However, if the $J = 30 - 29$ line had a compact, point-like, component at the position of IRC+10216 with a similar peak brightness to the SW clump (as is the case with $J=28-27$), we would expect to detect that even at the limited sensitivity of these data. Due to the similar excitation conditions of the $J = 30 - 29$ and $J=28-27$ lines, we take this as further evidence of the central peak in Fig.\ \ref{fig:tripfig1}a) being contributed by \ce{CH2NH}, and not \ce{HC3N}. Nevertheless, because of the observational limitations of this ALMA project (2011.0.00229.S), we do not include this line in our radiative transfer analysis.

Figure \ref{fig:co_contours} shows the observed \ce{HC3N} lines as contours overlaid on the image of \ce{^{13}CO} $J = 3 - 2$ at $\nu_{\mathrm{rest}}=330\,587.97$\,MHz created from the Band 7 ALMA observations. Using a \ce{CO} abundance of $6\times10^{-4}$ \citep{DeBeck2012_IRCphysCO,Agundez2012_IRCinnermols} and a \ce{^{12}C}/\ce{^{13}C} ratio of 48 \citep{Knapp1985_12co13coCSEs}, we adopt a constant \ce{^{13}CO} abundance of $1.3\times10^{-5}$ with respect to \ce{H2} throughout the regions of IRC+10216 we are investigating. Therefore, it is a useful indicator of the density structure. Keeping in mind the spatial filtering effects due to the lack of very short antenna spacings in these data, we see that the structures traced by \ce{HC3N} are spatially coincident with density enhancements resulting from non-isotropic mass-loss. The clump SW of IRC+10216 which is seen in all three observed \ce{HC3N} lines is also apparent here as the brightest area of \ce{^{13}CO} emission away from the central point source (see Fig.\ \ref{fig:co_contours}). Between the bright arcs we see several areas of negative flux, which are image artifacts resulting from resolved out emission and not indicative of actual absorption. When compared with images in \cite{Agundez2015_CH3CNmaps_IRC+10216}, we find that both \ce{HC3N} and \ce{^{13}CO} trace similar compact structures to \ce{CH3CN} ($J_K=14_3-13_3)$, including the SW clump. 

\subsection{Radiative Transfer Modeling} 
\label{RTmodels}
\begin{figure}[t!]
    \centering
    \includegraphics[width=\linewidth]{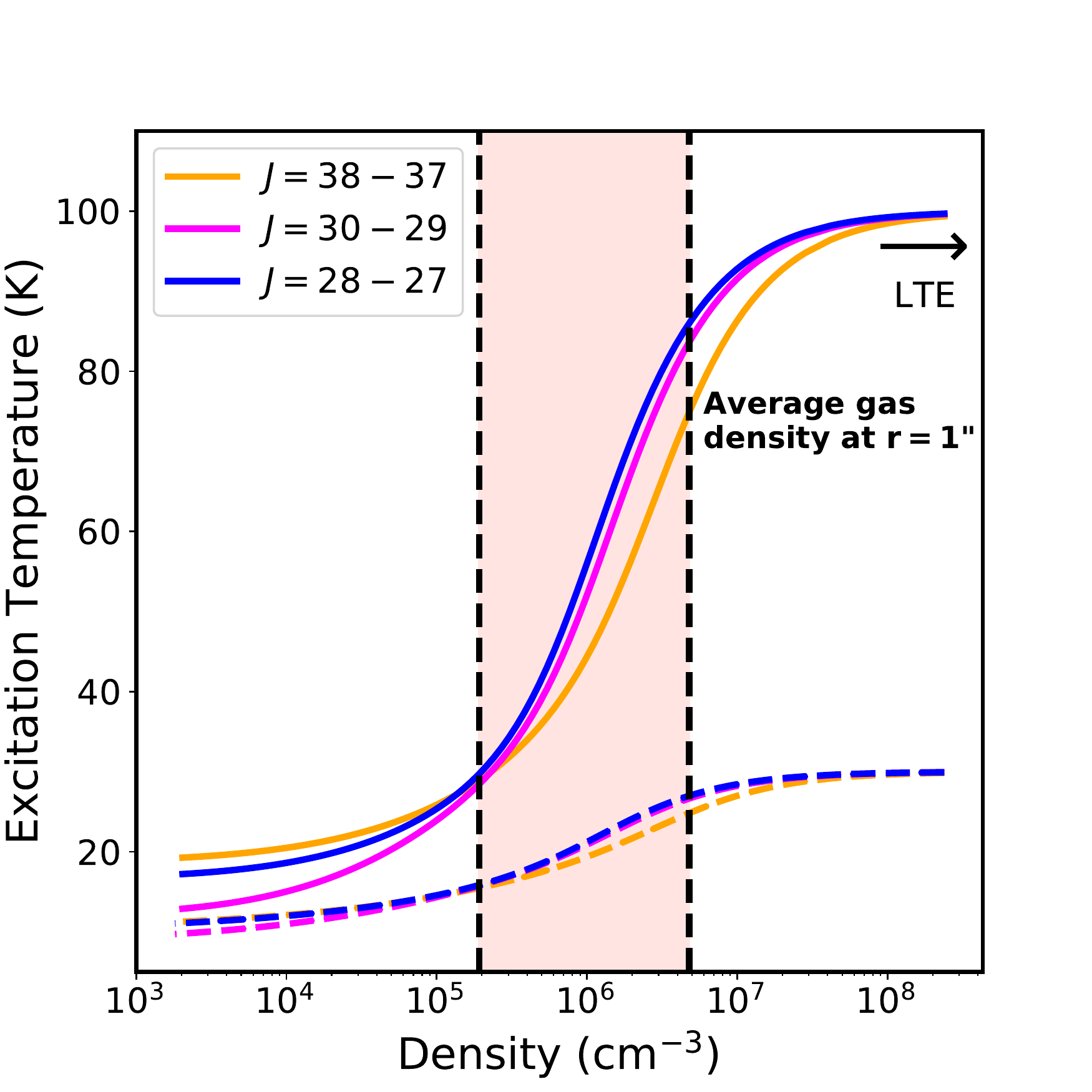}
    \caption{RADEX models of the three observed transitions of \ce{HC3N} run with for a range of densities at $T_k=30$\,K (dashed lines) and $T_k=100$\,K (solid lines). The shaded region denotes the average gas densities at radii where we observe these lines toward IRC+10216 (between 1\arcsec and 5\arcsec).}
    \label{fig:crit_dens}
\end{figure}

In order to draw conclusions regarding the abundance of \ce{HC3N}, we must first consider the excitation physics of this molecule in the inner envelope of IRC+10216. In many cases, Local Thermodynamic Equilibrium (LTE) is a good approximation for chemical studies of CSEs due to their high central densities. However, the rotational transitions observed here have comparatively large Einstein $A_{ij}$ coefficients ($10^{-2.9}-10^{-2.5}$\,s$^{-1}$), so it is possible that they are sub-thermally excited even at arcsecond scales. In Figure \ref{fig:crit_dens}, we show the predicted excitation temperatures of the three \ce{HC3N} transitions as a function of gas density for two different kinetic temperatures, generated using the RADEX radiative transfer software \citep{VanDerTak2007_RADEX}. Using the 1D density law from (Eq.\ \ref{eq:gas_dens} in \citeauthor{Agundez2012_IRCinnermols}~\citeyear{Agundez2012_IRCinnermols}) along with a mass-loss rate of $2.7\times10^{-5}$\,\msunyr\, \citep{Guelin2018_IRC_3D_Morph}, we also show the expected average gas densities at the radii where most of the emission is seen (between 1\arcsec and 5\arcsec). From this, it is clear that LTE could only be assumed if densities were higher than $10^8$\,\cmnegthree, and the observed gas density is below the critical densities of all three \ce{HC3N} transitions in the mapped regions. Futhermore, pumping to vibrational states via thermal infrared (IR) photons from dust grains has been shown to be an important excitation mechanism in the envelope of IRC+10216 \citep{Deguchi1984_hc5nIRpump,Cernicharo2014_IRC_timevar,Agundez2017_IRC_carbonchains,Keady1993_IRC_IR_lines,Fonfria2008_c2h2_hcn}, so non-LTE calculations are necessary to take all of these factors into account. 

Therefore, we adopt a forward modeling approach in predicting the expected brightness distributions and determining the abundance of \ce{HC3N} needed to reproduce the maps in Fig.\ \ref{fig:tripfig1}. We utilize the LIne Modeling Engine (LIME)  to simulate the level populations of \ce{HC3N}, solve the radiative transfer equation, and produce synthetic image cubes \citep{Brinch2010_lime}. LIME uses density-weighted grid points distributed in a computational space using Delaunay triangulation. Populations are solved based on the local conditions of the corresponding Voronoi cells, as well as the incident radiation from adjacent cells. For all our calculations, we use 50\,000 grid points in a computational volume with radius 130\,000\,AU (1000\arcsec). Though LIME performs these calculations in 3D space, we use a 1D approximation for the physical structure of IRC+10216, since the 3D asymmetries apparent in Figures \ref{fig:tripfig1} and \ref{fig:co_contours} are not presently constrained in their local physical conditions and require hydrodynamical simulations that are outside the scope of this work. In general, we adopt the same model for the CSE of IRC+10216 as \cite{Agundez2012_IRCinnermols}. This includes the stellar parameters, velocity structure, gas and dust temperature profiles, and density gradient. The only difference in our physical model is the use of a slightly larger mass-loss rate of $2.7\times10^{-5}$\,\msunyr\,, as this value was measured by \cite{Guelin2018_IRC_3D_Morph} specifically using \ce{CO} emission at radii within 15\arcsec, meaning it is better suited for this study of warm \ce{HC3N} emission. The gas density in the extended envelope is given by 
\begin{equation}
\label{eq:gas_dens}
    n(r) = \frac{\Dot{M}}{4\pi r^2 v_{\infty} m_g}
\end{equation}
where $\Dot{M}$ is the mass-loss rate, $r$ is the radius, $v_{\infty}$ is the expansion velocity, and $m_g$ is the average gas particle mass taken as 2.3\,amu considering \ce{H2}, \ce{He}, and\ce{CO} \citep{Agundez2012_IRCinnermols}. So a change in $\Dot{M}$ has a significant effect on molecular volume densities and corresponding line intensities. This should be kept in mind when comparing our results with previous chemical studies of IRC+10216 that use the more typical value of $2.0\times10^{-5}$\,\msunyr\, \citep{Agundez2017_IRC_carbonchains,Agundez2015_CH3CNmaps_IRC+10216,QL2017_NaCN_IRC}.

For spectroscopic data, the Leiden Atomic and Molecular Database (LAMDA) provides collisional rates of \ce{^{13}CO} and \ce{HC3N} with ortho- and para- \ce{H2} \citep{vanderTak2020_lamda}. The ortho/para \ce{H2} ratio is assumed to be 3 at all locations in the envelope. For \ce{HC3N}, the collisional rates measured by \cite{Faure_HC3Ncol} only go up to the $J=37$ rotational state, so we extrapolated these to the $J=45$ level and use line strengths from CDMS for the added radiative transitions \citep{2005JMoSt.742..215M}. The $J=45$ level was chosen to account for population in rotational states above $J=38$ while keeping the computational load low. We note that the choice of this maximum level had little effect on the predictions, as we found that running the model with maximum $J=40$ and $J=50$ yielded nearly identical results.

In addition, we included the first excited states of the $\nu_5$ and $\nu_6$ vibrational modes at 663\,cm$^{-1}$ and 499\,cm$^{-1}$ \citep{Jolly2007_HC3N_v5v6} using radiative rates from the HITRAN database \citep{Gordon2022_HITRAN}. The effect of including these states is similar to the result of \cite{Massalkhi2019_CS_SiO_SiS_Cstars}, where emission is shifted to larger radii as cooler molecules could be pumped to higher $J$ values by first being radiatively excited into these vibrational states then subsequently de-exciting through the P-branch ($\Delta J=+1$) \citep{Costagliola2010_HC3Nvib_NGC4418}. Collisional transitions within vibrational states were not considered, as the time scales of vibrational de-excitation (${\sim}10^{-4}$\,s) are several orders of magnitude shorter than the expected collisional excitation rates (${\sim}10^{4}$\,s). To model thermal emission from dust grains, we use the opacity law of \cite{Ossenkopf1994_dustopac}.

After obtaining ray-traced simulated image cubes from LIME, we test two separate methods of post-processing before comparing with observations. The first involves a simple image plane continuum subtraction followed by convolution of the channel maps with the synthesized beams in Fig.\ \ref{fig:tripfig1}. For the second method, we utilize the Python module \texttt{vis\_sample}\footnote{\texttt{vis\_sample} is publicly available under the MIT license at \url{https://github.com/AstroChem/vis_sample} and described in further detail in \citep{Loomis2018_MF_interfero}} to simulate the incomplete sampling of sky brightnesses using input \textit{uv} grid points directly from the observation sets. The results are simulated visibilities that were then continuum subtracted and imaged in the same manner as their respective observations (described in Section \ref{observations}). The estimated flux error in the simulated image is 5\%, based on the converged signal-to-noise of the individual grid points. Because the input model is spherically symmetric, we take the azimuthal average of the LIME-generated channel map at the systemic velocity of IRC+10216 to compare the modeled emission with ALMA images.

The result of both methods when applied to the $J=3-2$ line of \ce{^{13}CO}, for which the abundance is known \textit{a priori}, is shown in Fig.\ \ref{fig:azprofs}a. The convolved LIME model of this line overpredicts the brightness, especially in the extended layers of IRC+10216. In contrast, the visibility sampled and reconstructed image reproduces the peak brightness to within 5\%, and predicts the much lower average brightness of this line due to spatial filtering. From this it is clear that the latter method better accounts for the lack of data from short baselines and the resulting flux loss at large spatial scales, so we employ the visibility sampling routine for the case of \ce{HC3N} as well.

\subsubsection{Results for \ce{HC3N}}

\begin{figure*}[h!]
    \centering
    \includegraphics[width=\linewidth]{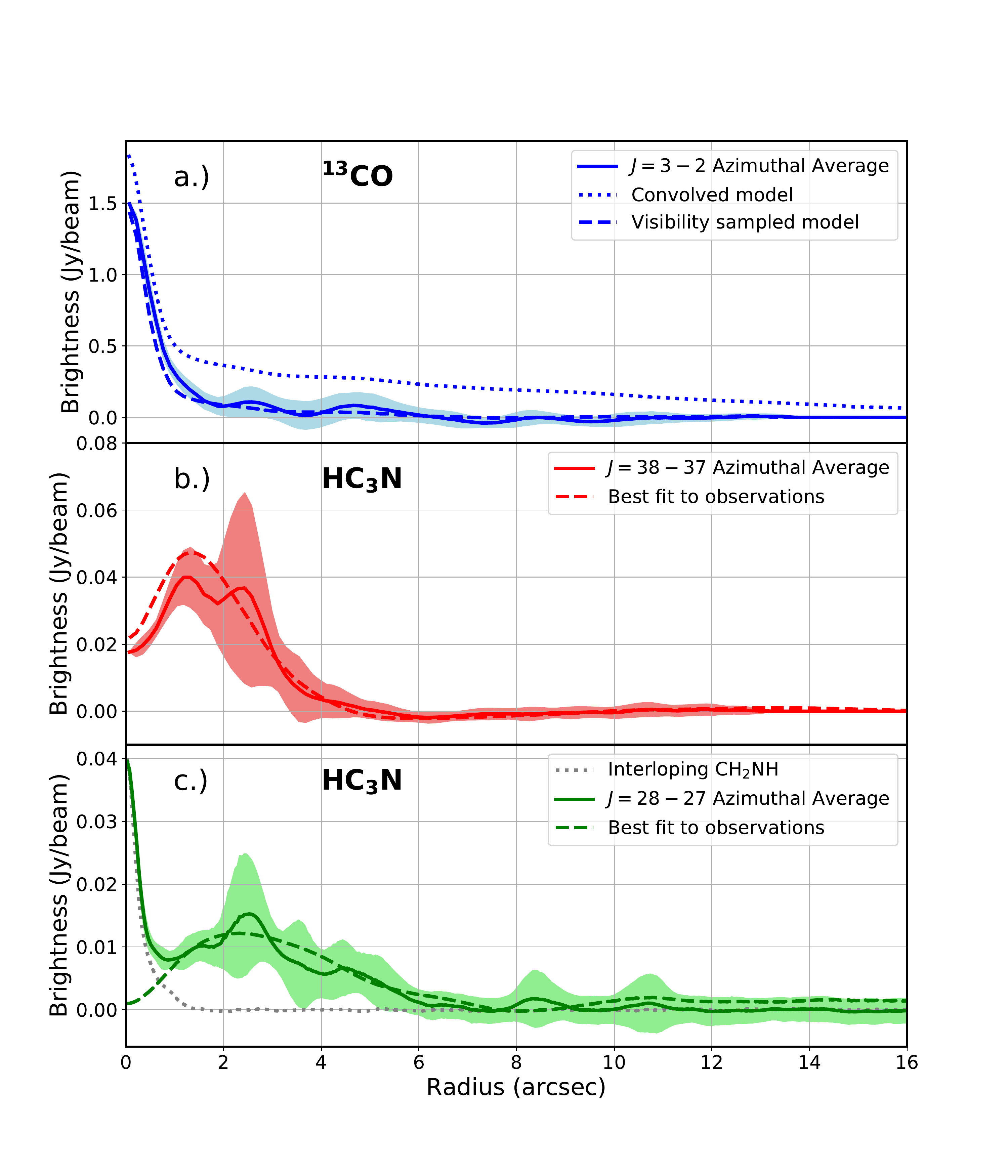}
    \caption{Observed (solid lines) and LIME simulated (dashed lines) brightness profiles of \ce{^{13}CO}, \ce{HC3N} $J=38-37$, and \ce{HC3N} $J=28-27$ from top to bottom. Profiles were obtained by taking the azimuthal average of the channel map at the systemic velocity of IRC+10216 centered on the position of the AGB star. Shaded regions denote the standard deviation of the surface brightness in each radial bin. In the top panel, the profile is also shown when a convolution with the synthesized beam of observations is used instead of visibility sampling. In the bottom panel, the expected contaminating emission from \ce{CH2NH} is shown as well (dotted line).}
    \label{fig:azprofs}
\end{figure*}

\begin{figure}[t!]
    \centering
    \includegraphics[width=\linewidth]{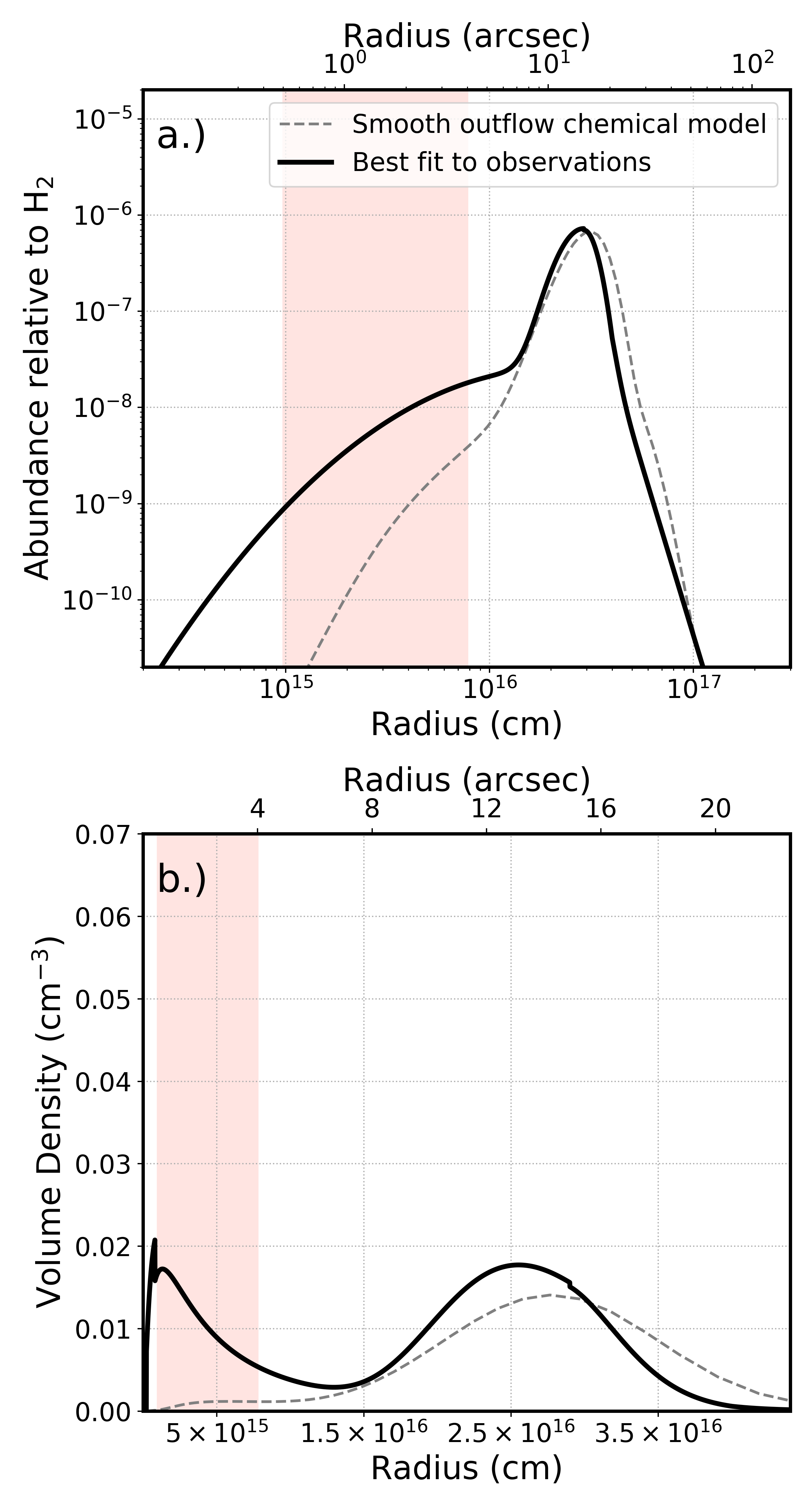}
    \caption{Radial abundance with respect to \ce{H2} (top) and volume density (bottom) profiles of \ce{HC3N} in the envelope of IRC+10216. Solid black line represents the best fit profile resulting from the visibility sampled LIME radiative transfer models, while the dashed gray line is the prediction of a simple kinetic chemistry model (described in Section \ref{chem_models}). The shaded red region represents the radii where the \ce{HC3N} abundance is best sampled by the observations.}
    \label{fig:abund}
\end{figure}

We use the above forward-modeling routine to simulate the $J=28-27$ and $J=38-37$ lines of \ce{HC3N}, now varying the abundance to reproduce the observed radial brightness distribution. Primarily, we modify it at radii less than 4\arcsec, since this is where we see the bulk of emission (shaded regions in Fig.\ \ref{fig:abund}); however, we also adjust the peak abundance to reflect the observed lack of emission at radii larger than 8\arcsec. We start with a function similar to the modeled result in \cite{Agundez2017_IRC_carbonchains}, with a peak abundance of 1$\times10^{-6}$ at $r=3\times10^{16}$\,cm that steadily decreases inwards, and drops off steeply outwards. We then adjust the magnitude shape to match the distribution and intensity, by eye, of azimuthally averaged emission for both \ce{HC3N} lines.

The resulting simulated intensity profiles are shown in Figures \ref{fig:azprofs}b.\ and \ref{fig:azprofs}c. The corresponding best-fit abundance profile is shown in Fig.\ \ref{fig:abund}a., and the volume density is shown in Fig.\ \ref{fig:abund}b. The resulting modeled intensity curves are in good agreement with the observed azimuthal averages at the radii where we see the bulk of \ce{HC3N} emission. There is a slight disagreement in the regions between 8\arcsec--14\arcsec where the average predicted $J=28-27$ emission increases 2\,\mjb; however, this is very close to the rms of the observations, so it is not expected that we would detect consistent emission at this level. We do detect some clumps with a peak abundance of 25\,\mjb\ in these regions, as seen outside the yellow reference circle to the SW and NW in Fig.\ \ref{fig:tripfig1}a., but no extended component is found.

In the inner layers where the \ce{HC3N} emission is primarily seen, an average abundance of ${\sim}10^{-8}$ is needed, and it must be decreasing inward to avoid centrally peaked channel maps. In typical fashion for characterizing the radial abundance distributions of molecules forming in outer regions of CSEs \citep[e.g.\ ][]{Agundez2015_CH3CNmaps_IRC+10216}, we parameterize this with an offset Gaussian in log radius:

\begin{equation}
    f\left(r\right) = f\left(r_p\right)\mathrm{exp}\left(-\left(\frac{\log r - \log r_p}{\Delta \log r}\right)^2\right)
\end{equation}
where $f$ is the fractional abundance with respect to \ce{H2}, $r_p$ is the peak radius, and $\Delta \log r$ adjusts the width of the feature. For the retrieved profile, we obtain $f\left(r_p\right)=2.5\times 10^{-8}$, $r_p=2\times10^{16}$\,cm, and $\Delta \log r=1.65$. This is the main function we use to describe the \ce{HC3N} abundance in the regions of IRC+10216 probed by our observations, and an additional narrow component peaking at $r=3\times10^{16}$\,cm with $f=7\times 10^{-7}$ is included to reflect radii larger than $10^{16}$\,cm. This is smaller than the peak derived in \citet{Agundez2017_IRC_carbonchains} by a factor of 1.4, but this inconsistency is expected since a smaller mass-loss rate is used in that work. The retrieved abundance profile is shown in Fig.\ 7a.\ along with the results of a chemical model for IRC+10216, which is described in Section \ref{chem_models}. 

\section{Discussion}
\label{disc}
When compared with the results of our own chemical model as well as the abundances in \cite{Agundez2017_IRC_carbonchains}, the measured \ce{HC3N} abundances demonstrated in Fig.\ \ref{fig:abund} is about one order of magnitude larger than predictions of kinetic chemistry at radii less than $10^{16}$\,cm. The disparity with \citet{Agundez2017_IRC_carbonchains} is actually greater than that since a lower mass-loss rate was adopted in that work, which will naturally yield higher abundances necessary to reproduce a given observation. 

One factor that could potentially affect abundance calculations in the envelope of IRC+10216 is the variability of rotational molecular lines with stellar phase. \citet{Pardo2018_IRC_molvariable} observed periodicity in 3\,mm radical, polyyne, and cyanopolyyne species (including \ce{HC3N}) that can be either in or out of phase with the ${\sim}$640 day periodicity of the AGB star as seen in the infrared. \citet{He2017_IRC_line_monitor} observed the same behavior in higher frequency lines, including the $J=28-27$ blend mapped in this work. We do not expect this to have a dramatic effect on our results, mainly because the amplitude of these oscillations are relatively small (ranging from 8\%--15\% of the average integrated line flux), but also because the combined data we use for the $J=28-27$ line was taken, and thereby averaged, over the course of five months, almost a quarter of the full period. Nevertheless, if we extrapolate the light curve from \citet{He2017_IRC_line_monitor} for the periodicity to the dates of the ALMA observations (Nov.\ 2019 -- Mar.\ 2020), we find that the map in Fig.\ \ref{fig:tripfig1}a represents phases near an expected \textit{minimum} of this line. The $J=38-37$ image was made from a single observation in May 2018, meaning it would instead be expected to near the average brightness if it follows the same function. Once again, however, this is likely a very minor effect on the retrieved abundances since the amplitude of variation is small, especially when taken in comparison to the brightness variation from the asymmetric 3D nature of the outflow (illustrated by the shaded curves in Fig.\ \ref{fig:azprofs}), which is the primary source of uncertainty in our analysis.

With this in mind, our results indicate that some process is enhancing the photochemical production of \ce{HC3N} in the envelope of IRC+10216. Therefore, our observations of \ce{HC3N} in the inner layers of IRC+10216 follow a pattern of previous results for product molecules in this envelope, namely the studies of \ce{CH3CN}, \ce{NaCN}, \ce{H2O}, and \ce{C4H2} \citep{Agundez2015_CH3CNmaps_IRC+10216,QL2017_NaCN_IRC,Agundez2010_H2O_CSE_photo,Fonfria2018_IRC_C4H2}. The former is driven by the radiative association of photoionized \ce{CH3+} with \ce{HCN}, while \ce{NaCN} is thought to form via a reaction between \ce{NaCl} and \ce{CN} \citep{Agundez2008_IRC_CH2CNmore,Petrie1996_MCN_form}. \ce{H2O} and \ce{C4H2} have formation routes through reactions with the photodissociation products \ce{C2H} and neutral \ce{O}. \ce{HC3N} formation in the extended CSE is likewise governed by the following reactions:
\begin{equation}
\mathrm{HCN} + h\nu \rightarrow \mathrm{CN} + \mathrm{H}
\end{equation}
\begin{equation}
\mathrm{C_2H_2} + \mathrm{CN} \rightarrow \mathrm{HC_3N} + \mathrm{H}
\end{equation}
In common between all these reaction pathways is the need for a source of UV photons to provide the necessary reactants. The anomalous overabundances for all four of these species thus suggest a larger flux of UV radiation than expected in the inner envelope. \cite{Agundez2015_CH3CNmaps_IRC+10216} investigated the possibility that \ce{CH3CN} abundances result from the asymmetric clumpiness of the outflow allowing ionizing photons from the ISRF to penetrate farther into the outflow than the smooth spherical case; however, this was also not enough to reproduce their observations. Here, we explore an alternate scenario where the missing UV is provided internally from a binary companion star orbiting IRC+10216 at a close distance.

Evidence of such a companion existing nearby IRC+10216 has grown over recent years, primarily through spatial studies of \ce{CO} and its isotopologues. \cite{cernicharo2015_IRC_MLRhist} and \cite{Guelin1993_MgNC_IRCmap} note the 16\arcsec (${\sim}$700\,yr) spacing of the outermost bright gas shells which is neither consistent with the approximate timescales of thermal pulses (${\sim}10^4$\,yr) nor the pulsations exhibited by AGB stars (${\sim}1$\,yr). These authors suggest that the shells could instead be explained by a companion on an elliptical orbit periodically affecting the mass-loss rate of IRC+10216 during periastron. \cite{Guelin2018_IRC_3D_Morph} found that the spacing of these shells decreases moving inwards, indicating mass transfer or additional dynamics occurring. In addition, \cite{Decin2015_IRC_spiral} detected signatures of spiral arm structure at arcsecond scales which are consistent with a main-sequence 1\msun\, companion orbiting at ${\sim}20$\,AU with an orbital period of 55\,yr. Lastly, \cite{Kim2015_IRC_companion_HST} and \cite{Kim2021_IRC_HSTmonitor2} detected a secondary point source of optical light about 0.5\arcsec east of IRC+10216 using HST monitoring. Based on its spectral index, they argue that the object could be a binary companion of spectral type M1, but this result has yet to be confirmed. In proposed scenarios of \citet{Kim2015_IRC_companion_HST} and \citet{Decin2015_IRC_spiral}, the binary separation is less than the minimum beam size of our observations, and in the case of \citet{Guelin2018_IRC_3D_Morph}, the proposed separation would be ${\sim}1"$. So we expect the region of the possible companion to be unresolved, or at most very weakly resolved, from the central AGB star in our observations.

While the exact nature of the orbital system IRC+10216 appears to be a part of is still not well-characterized due to uncertainties in the above works, a binary scenario seems very likely. Although IRC+10216 itself is likely far too red to produce enough UV flux to drive photochemistry in the inner envelope, a companion with a higher surface temperature could accomplish this. To test this scheme in comparison to our measurements of \ce{HC3N}, we utilize a detailed chemical model of IRC+10216.  

\subsection{Chemical modelling}
\label{chem_models}
The chemical kinetics model is that of \citet{VandeSande2022}, which includes the effect of stellar companion UV photons on the gas-phase chemistry throughout the outflow. 
Several assumptions go into the model to reduce the complexity of the problem. The one-dimensional model does not include dust formation and growth, but assumes dust is present from a radius \Rdust\ onwards.
The companion is assumed to be at the center of the outflow and lie within the dust-free region before \Rdust. 
Considering the scale of the outflow, we consider the effect of misplacing the companion within the dust-free region is negligible.
The photon flux of the companion is approximated by blackbody radiation and is set by the stellar radius, $R_*$\, and blackbody temperature, \Tcomp.
The one-dimensional nature of the model implies that orbital motion cannot be taken into account. 
Hence, the companion's radiation field is assumed to be always present. 
This assumption is explored further in Appendix \ref{geom}.
The model assumes the companion to be one of three types: a red dwarf at 4000\,K, a solar-like star at 6000\,K or a white dwarf at 10\,000\,K.

The effects of a clumpy outflow on the chemistry are taken into account using the porosity formalism, as described by \citet{VandeSande2018}.
The formalism divides the outflow in a stochastic two-component medium, consisting of a rarified interclump component and an overdense clumped component, and allows us to include the influence of a clumpy outflow on the penetration of UV photons along with the effect of local overdensities.
Dust and gas are assumed to be well-mixed in the interclump and clumped components.
The specific locations of the clumps cannot be specified. 
Rather, the clumpiness of the outflow is determined by three general parameters: the interclump density contrast, \fic\, which describes the distribution of material between the clump and interclump component, the volume filling factor, \fvol\, which sets fraction of the total outflow volume occupied by clumps, and the size of the clumps at the stellar surface, $l_*$.
As volume filling factor is assumed to be constant throughout the outflow and the clumps are assumed to be mass conserving, the clumps expand as they move away from the AGB star.

\begin{figure}[t!]
    \centering
    \includegraphics[width=\linewidth]{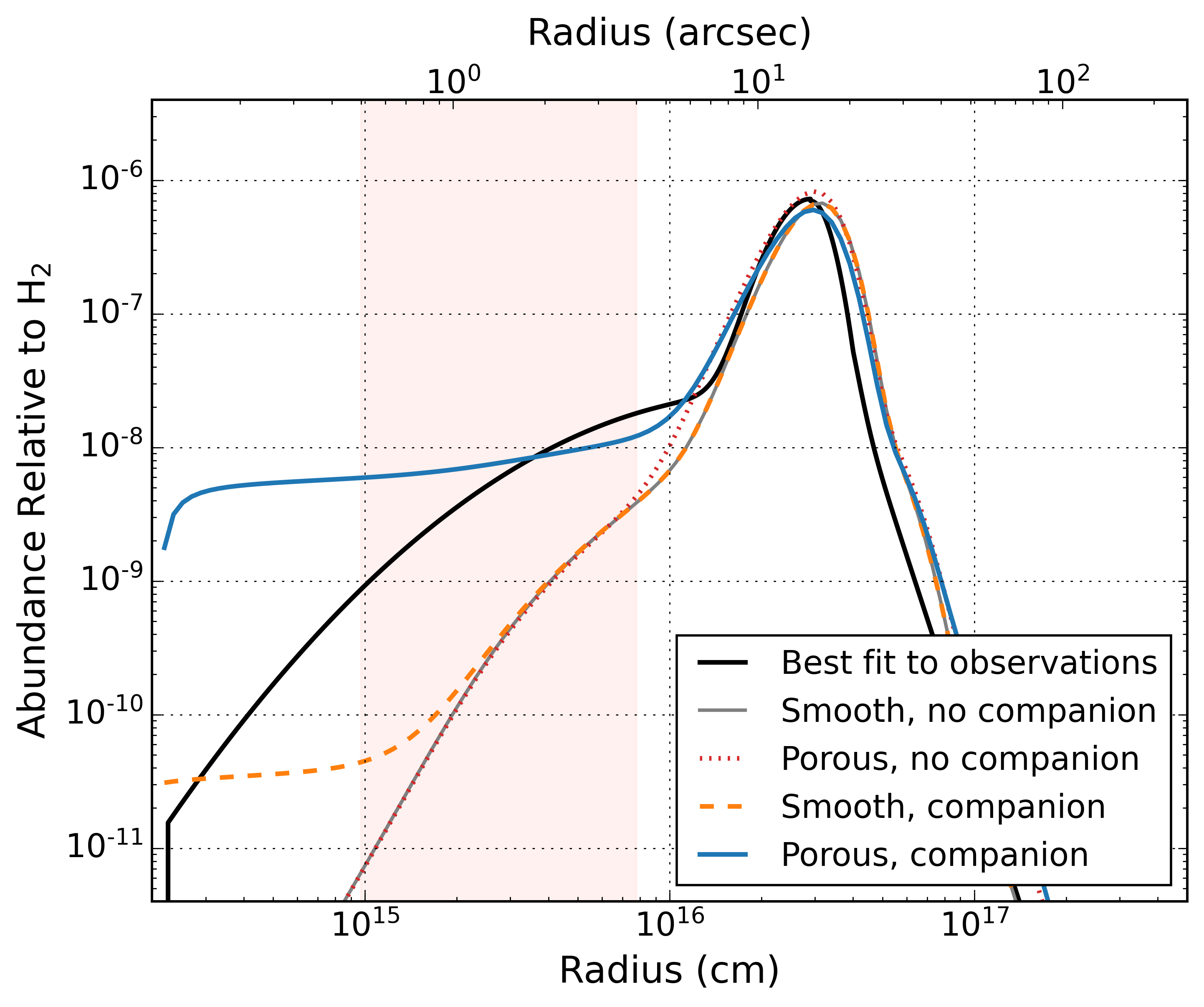}
    \caption{The retrieved \ce{HC3N} abundance (black) together with the chemical modelling results. Gray, solid: smooth outflow without a stellar companion. Red, dotted: porous outflow (characterised by \fic\ = 0.40, \fvol\ = 0.50, and $l_* = 2 \times 10^{13}$ cm) without a stellar companion. Orange, dashed: smooth outflow with a solar-like companion and dust condensation radius \Rdust\ = 5 R$_*$. Blue, solid: porous outflow with a solar-like companion and  \Rdust\ = 5 R$_*$. The shaded red region represents the radii where the \ce{HC3N} abundance is best sampled by the observations.}
    \label{fig:chemicalmodel}
\end{figure}

The values of the three porosity parameters are observationally constrained. 
The shells and their clumps were resolved in the VLA survey of \citet{Keller2017}.
At a distance of about 24\arcsec from the star, the angular sizes of the clumps were found to range from 2.0\arcsec $\times$ 3.0\arcsec up to 3.0\arcsec $\times$ 7.7\arcsec. 
Assuming a distance of 130 pc, this corresponds to an upper limit on the physical sizes of the clumps of $3.9 - 15.6 \times 10^{15}$ cm at a distance of $\sim 4.6\times 10^{16}$ cm from the star.
This constrains the value of $l_*$ to $2 - 4 \times 10^{13}$ cm. 
The density contrast between the shell and intershell region was found to lie between 1.5 and 7.6 at a radial distance of $\sim$ 15\arcsec\ from the star.
This is consistent with the shell-intershell brightness contrast of $\geq$ 3 retrieved by \citet{Cernicharo2015}. 
Assuming this density contrast can be applied to the clumped and interclump components, the values of \fic\ and \fvol\ are linked following
\begin{equation}
    \frac{\rho_\mathrm{cl}}{\rho_\mathrm{ic}} = \frac{1 - (1 - f_\mathrm{fvol}) f_\mathrm{ic}}{f_\mathrm{fvol}\ f_\mathrm{ic}},
\end{equation}
with $\rho_\mathrm{cl}$ and $\rho_\mathrm{ic}$ the clumped and interclump densities, respectively. 
Lastly, we assume that dust grains are purely absorbing and therefore neglect the scattering of UV photons. \citet{Jones2013} showed that for small hydrocarbon dust grains, that provide the extinction at UV wavelengths, their albedo is essentially zero at these wavelengths. 
Such small grains are thought to be present in the inner envelope \citep{Fonfria2022}.  
Although the Jones et al. models are strictly for interstellar dust, their results are likely to hold for IRC+10216; thus scattering can largely be ignored in the inner regions of the outflow.

The physical parameters adopted for the outflow of IRC+10216, together with those derived for our best-fit chemical model, are listed in Table \ref{table:physpar-chemmodel}. 
The parent species used are those of \citet{Agundez2012_IRCinnermols}, but with the initial abundance of \ce{C2H2} and HCN decreased by a factor two. This was done to ensure agreement with the results from \citet{Fonfria2008_c2h2_hcn}, who observed mid-infrared vibrational lines of these molecules coming from the innermost regions of the envelope. Halving the initial \ce{C2H2} and HCN lies within the error of the abundances they retrieve for the region closest to the star, from the stellar photosphere up to 5.2 R$_*$, which is where our chemical model starts.

We find that to reproduce the increase in HC$_3$N in the inner region, both clumpiness and a companion are necessary.
The increase is best reproduced assuming a solar-like companion and a dust condensation radius of 5 R$_*$ together with an interclump density contrast \fic\ = 0.40.
Figure \ref{fig:chemicalmodel} shows the retrieved abundance profile along with chemical model results, for smooth and porous outflows both with and without a solar-like companion. 
The value of \Rdust\ corresponds to the results of \citet{Fonfria2022}, who found gradual dust growth up to 5 R$_*$.
While the chemical model does not include dust growth, we find that this value of \Rdust\ best reproduced the retrieved abundance profile.
The values of \fvol\ and $l_*$ do not have a large influence on the shape of the abundance profile and are constrained by the observational results of \citet{Keller2017}. 

Although a mass-loss rate of $2.7 \times 10^{-5}$\,\msun\ yr$^{-1}$ was used for the radiative transfer model to describe the envelope density within a radius of 15\arcsec \citep{Guelin2018_IRC_3D_Morph}, a lower one is needed for the chemistry model since the entire extended envelope contributes to the extinction of UV photons, and lower average mass-loss rates have been obtained when considering radii out to 300\arcsec \citep{Decin2011_IRCshells,Teyssier2006_AGB_CO_obs,cernicharo2015_IRC_MLRhist}. Therefore, to better fit the attenuation of the extended molecular shell, the mass-loss rate was changed from $2.7 \times 10^{-5}$\,\msun\ yr$^{-1}$, as retrieved by \citet{Guelin2018_IRC_3D_Morph} to $1.8 \times 10^{-5}$\msun\ yr$^{-1}$.

While the porous model with a solar-like companion shows an increased HC$_3$N abundance before the molecular shell at $\sim$ 15\arcsec, the gentle increase in the retrieved abundance cannot be reproduced.
This is because the model cannot take the gradual dust growth in the inner wind into account \citep{Fonfria2022} and starts only after the dust condensation radius, leading to an abrupt start of photochemistry in the inner wind and subsequently a sharp increase in HC$_3$N abundance. 
Additionally, the porosity formalism provides only a one-dimensional approximation of the three-dimensional clumpy substructure seen in IRC+10216's outflow.
Nevertheless, it is once again clear that a classical chemical model of smooth outflow without a stellar companion does not suffice.

\begin{deluxetable}{ll}
\label{table:physpar-chemmodel}
	\tablecaption{Physical parameters of the chemical model}
	\tablehead{\multicolumn{2}{c}{Parameters adopted from \citet{Agundez2012}}}
    \startdata
	Outflow velocity, $v_\infty$ 	& 14.5 km s$^{-1}$ \\
    Stellar radius, $R_*$             & 4 $\times 10^{13}$ cm \\
    Stellar temperature, $T_*$        & 2330 K \\
    Exponent $T(r)$, $\epsilon$                    & 0.5 \\
	\multicolumn{2}{c}{Modelling results}\\ \hline
    Mass-loss rate, $\dot{M}$        	&    $1.8 \times 10^{-5}$  $\mathrm{M}_\odot\ \mathrm{yr}^{-1}$  \\
    Interclump density contrast, \fic\			& 0.50 \\
    Clump volume filling factor, \fvol\		& 0.25 - 0.80 \\
    Companion temperature., \Tcomp, 					& 6000 K \\
    Companion radius, \Rcomp,					& $8.14\times 10^{10}$ cm \\
    Onset of dust extinction, $R_\mathrm{dust}$		& 5 $R_*$ \\
    Start of the model		& $1.025 \times R_\mathrm{dust}$ \\
    \enddata

\end{deluxetable}

\section{Conclusions}
We have presented an investigation of warm \ce{HC3N} emission in the inner layers of IRC+10216 utilizing multiple archival ALMA observations. Mapping the $J=28-27$, $J=30-29$, and $J=38-37$ lines of this molecule reveals the most compact distribution ($r<10^{16}$\,cm) of a cyanopolyyne molecule presently detected toward the outflow of an evolved star, with emission localized to radii where \ce{CH3CN}, \ce{NaCN}, \ce{H2O}, and \ce{C4H2} were previously observed \citep{QL2017_NaCN_IRC,Agundez2015_CH3CNmaps_IRC+10216,Agundez2010_H2O_CSE_photo,Fonfria2018_IRC_C4H2}. The emitting region is almost completely distinct from previous works that imaged lower excitation lines of \ce{HC3N}, allowing for an in-depth characterization of this molecule throughout the CSE. Using a 1D model of the physical conditions surrounding IRC+10216, we used visibility sampled LIME simulations to derive an average abundance of $10^{-8}$ with respect to \ce{H2} for \ce{HC3N} at the radii probed by these observations. This is about five to ten times larger than predicted by simple chemical models of IRC+10216, suggesting that photochemistry is occurring more rapidly than expected in these warmer layers.

To explain this result, we investigated the possibility that the enhancement of \ce{HC3N} surrounding IRC+10216 is caused by an embedded binary companion emitting UV photons in the inner wind and driving photochemistry. Using known constraints on the physical environment and the abundances of parent molecules, we ran a specialized chemical kinetics network to model the radial \ce{HC3N} abundance profile for a variety of different companion scenarios. We find that invoking a solar-like companion and a dust condensation radius of 5\,R$_*$ produces a similar abundance feature in the inner wind to that observed, and conclude that this is an strong explanation for the anomalous abundances of product molecules observed at small radii toward IRC+10216.

This result is especially intriguing when considering the newfound prevalence of binary interactions in AGB outflows \citep[e.g.][]{Decin2020_atomium}. Since these systems appear to be common among nearby evolved stars, it could be that companions embedded in CSEs have a notable chemical affect on the composition of material returned to the ISM during the AGB and onward. In order to understand the full extent of this effect in IRC+10216, further millimeter and IR studies of photochemistry products (specifically their higher energy transitions) will be crucial. In addition, rigorous molecular line studies of C- and O-rich evolved stars that show signs of binary shaping could shed important light on this topic.

\section{Acknowledgements}
We thank the reviewer for their thorough and insightful comments on this manuscript. This paper makes use of the following ALMA data: ADS/JAO.ALMA\#2019.1.00507.S, ADS/JAO.ALMA \#2011.0.00229.S, ADS/JAO.ALMA\#2016.1.00251.S. ALMA is a partnership of ESO (representing its member states), NSF (USA) and NINS (Japan), together with NRC (Canada), MOST and ASIAA (Taiwan), and KASI (Republic of Korea), in cooperation with the Republic of Chile. The Joint ALMA Observatory is operated by ESO, AUI/NRAO and NAOJ. The National Radio Astronomy Observatory is a facility of the National Science Foundation operated under cooperative agreement by Associated Universities, Inc. Support for this work was provided to M.\,A.\,S.\, by the NSF
through the Grote Reber Fellowship Program
administered by Associated Universities, Inc./National
Radio Astronomy Observatory. M.\,A.\,S.\, also acknowledges additional support from the Virginia Space Grant Consortium.
M.\,V.\,d.\,S acknowledges support from the European Union’s Horizon 2020 research and innovation programme under the Marie Sk\l{}odowska-Curie grant agreement No 882991.
T.\,J.\,M gratefully acknowledges the receipt of a Leverhulme Emeritus Fellowship and the STFC for support under grant reference ST/P000312/1 and ST/T000198/1.

\bibliography{refs}
\bibliographystyle{aasjournal}

\appendix

\renewcommand{\thefigure}{A\arabic{figure}}
\renewcommand{\thetable}{A\arabic{table}}
\renewcommand{\theequation}{A\arabic{equation}}
\setcounter{figure}{0}
\setcounter{table}{0}
\setcounter{equation}{0}

\section{Effect of orbit geometry}
\label{geom}

\begin{figure}[h!]
    \centering
    \includegraphics[width=0.75\linewidth]{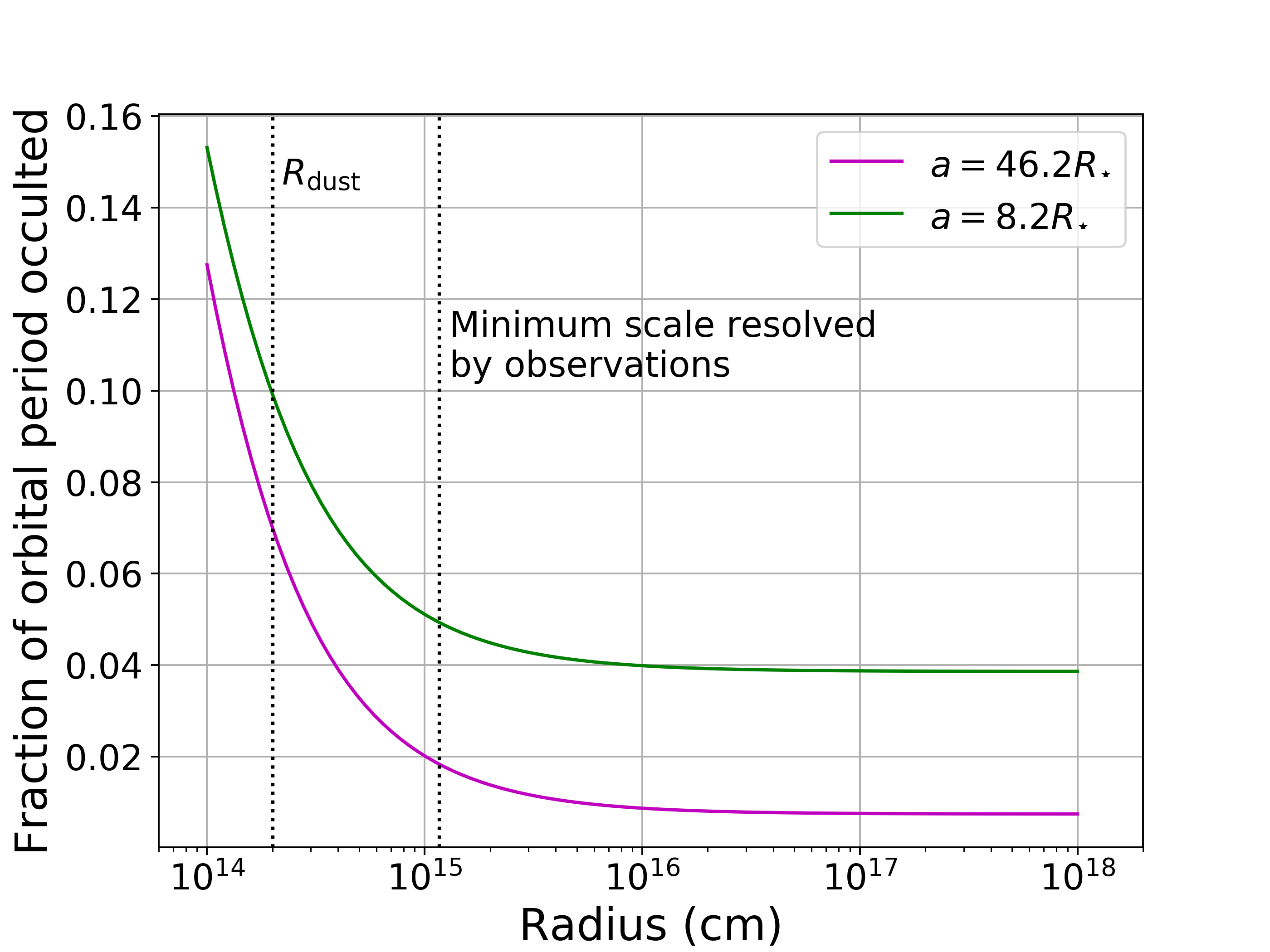}
    \caption{Occultation time of a companion star orbiting with IRC+10216 semi-major axis $a$ as seen by a point on the orbital plane as a function of radius. Magenta and green curves correspond to the semi-major axes proposed in \citet{cernicharo2015_IRC_MLRhist} and \citet{Decin2015_IRC_spiral}, respectively. }
    \label{fig:occult_time}
\end{figure}

The orbital motion of a stellar companion could induce chemical asymmetries in the outflow.
Because the chemical model is one-dimensional, orbital motion cannot be taken into account. 
Moreover, the companion star is assumed to be located at the center of the AGB star, resulting in a continuous irradiation of the outflow by the companion star.
\citet{VandeSande2022} argue that the limitations of the model are reasonable first-order approximations for a close-by companion with a short orbital period.
This assumption should be investigated in more detail for the IRC+10216 system because the tentative orbital solutions presented in \citet{cernicharo2015_IRC_MLRhist} and \citet{Decin2015_IRC_spiral} of respectively 700 and 55 yr cannot necessarily be considered sufficiently short. For a given point at radius $r$ in the outflow on the orbital plane of the binary system, assuming a circular orbit, the fraction of the period the companion spends behind the AGB star is given by 

\begin{equation}
    f \approx \frac{2a\arctan\left[\frac{r+a}{a}\frac{R_*}{r}\right]}{2\pi a}
\end{equation}

where $a$ is the semi-major axis and $R_*$ is the radius of the AGB star, provided $\bm{r\gg R_*}$. Using the empirical values from \citet{cernicharo2015_IRC_MLRhist} ($a=42.6\ R_*$) and \citet{Decin2015_IRC_spiral} ($a=8.2\ R_*$), we obtain the results shown in Fig.\ \ref{fig:occult_time}. In both cases, the perceived occultation time of the companion is less than 10\% of the total period at all radii larger than $R_{\mathrm{dust}}$. In the outer envelope, this value then drops to 4\% and 1\% for the long and short period solutions, respectively. From this, we would only expect time-dependent spatial chemistry differences to occur in the innermost regions of the envelope, and these scales are unfortunately not probed by our ALMA observations. Of course, other factors, including the varied dust extinction between the companion and the point as the companion orbits, the eccentricity, the presence of density enhancements, clumps, arcs, etc., do complicate the picture but as a first approximation it is clear that the companion photons are visible for a very large fraction of the orbit even in the long period solution from \citet{cernicharo2015_IRC_MLRhist}.

The other important detail here is the inclination of the orbit with respect to the observer. Since we are only mapping the \ce{HC3N} distribution at the systemic velocity, we would only expect chemical asymmetries to arise in regions opposite where the companion crosses the plane of the sky, which is a small fraction even for slight inclinations. Though the exact orbit is uncertain, \citet{Decin2015_IRC_spiral} proposed that the inclination is larger than $60^{\circ}$ based on observed \ce{^{13}CO} PV diagrams. An inclination this high would only cause chemical asymmetries (in the plane of the sky) to be observable if the envelope was observed during very specific points in the orbit, which is unlikely given the proposed period of 55 -- 700\,yr \citep{Decin2015_IRC_spiral,cernicharo2015_IRC_MLRhist}. Between this and the relatively short fractional occultation period, if a companion is indeed influencing \ce{HC3N} chemistry in the envelope of IRC+10216, we do not expect orbit-induced asymmetries to be prominent or detectable in the ALMA images presented in this work.

\end{document}